\documentclass[%
 reprint,
superscriptaddress,
showpacs,preprintnumbers,
 amsmath,amssymb,
 aps,
]{revtex4-1}

\usepackage{xcolor}
\usepackage{graphicx}% Include figure files
\usepackage{dcolumn}% Align table columns on decimal point
\usepackage{bm}% bold math
\usepackage[caption=false]{subfig}
\usepackage{fixmath}
\usepackage{epstopdf}

\newcommand\redout{\bgroup\markoverwith{\textcolor{red}{\rule[.5ex]{2pt}{0.4pt}}}\ULon}

\usepackage[colorlinks=true,citecolor=blue]{hyperref}
\hypersetup{colorlinks=true,citecolor=blue,linkcolor=red,urlcolor=blue}

\begin{document}
\title{Magnetoresistance in Noncentrosymmetric Two-dimensional Systems}
\author{Azadeh Faridi}
\email{azadeh.faridi@ipm.ir}
\affiliation{School of Physics, Institute for Research in Fundamental Sciences (IPM), Tehran 19395-5531, Iran}
\author{Reza Asgari}
\email{asgari@ipm.ir}
\affiliation{School of Physics, Institute for Research in Fundamental Sciences (IPM), Tehran 19395-5531, Iran}
\affiliation{School  of  Physics,  University  of  New  South  Wales,  Kensington,  NSW  2052,  Australia}

\date{\today}

\begin{abstract}
The valley-contrasting geometric features of electronic wave functions manifested in Berry curvature and orbital magnetic moment have profound consequences on magnetotransport properties in both three- and two-dimensional systems. Although the importance of employing beyond-relaxation-time-approximation methods and intervalley scatterings in collision integral has been confirmed in three dimensions, they have been widely overlooked in previous studies on two-dimensional multivalley systems. Here, we study the issue of weak-field magnetoresistance in two-dimensional multivalley systems with broken inversion symmetry. We provide an exact solution to the Boltzmann equation and demonstrate that the inclusion of in-scattering terms in the collision integral can change the sign of the magnetoresistance in the high-density regime. With an initial valley polarization, we also predict an orbital magnetic moment-induced intrinsic contribution to Hall conductivity in the time-reversal-broken situation, which is consistently negative, and in contrast to the anomalous Hall term, it does not depend on the polarization sign. Depending on which valley has the excess charge, our calculations show that a completely distinct behavior is exhibited in the magnetoresistance which can be considered as a valley-polarization probe in the experiment.      
\end{abstract}

%\pacs{71.10.-w, 71.18.+y, 73.21.−b}
\maketitle

\section{Introduction}
The history of magnetotransport studies goes back to the 19th century and the experiments of William Thomson (Lord Kelvin) on iron and nickel in the presence of a magnetic field~\cite{thomson1857xix}. Since then, the study of magnetotransport properties of materials has always been one of the leading and practical subjects of research in condensed matter physics revealing many significant features in materials ranging from the electronic structure via Shubnikov-de Haas oscillations to topologically nontrivial signatures in Dirac and Weyl semimetals~\cite{son2013chiral,burkov2014chiral,spivak2016magnetotransport}. 

Numerous intriguing and remarkable magnetotransport findings are closely related to the geometry of Bloch wave functions, which is primarily illustrated by Berry curvature and orbital magnetic moment (OMM)~\cite{xiao2010berry}. This can have more profound consequences in systems with broken inversion symmetry where the coupling between the valley degree of freedom with equal and oppositely oriented Berry curvature and OMM in two valleys can result in unique outcomes~\cite{xiao2007valley,mak2014valley,schaibley2016valleytronics}. The negative magnetoconductance in Weyl semimetals~\cite{knoll2020negative,sharma2020sign}, the chiral magnetic effect~\cite{ma2015chiral}, and the gyrotropic magnetic effect~\cite{zhong2016gyrotropic} are a few examples. An out-of-plane magnetic field can raise the valley degeneracy due to the valley-contrasting OMM~\cite{rostami2015valley}. While the valley-contrasting Berry curvature has long been considered in theoretical calculations of the magnetotransport in multivalley systems~\cite{xiao2010berry}, the equally important role of the OMM has currently seen a revived interest in both three- and two-dimensional (3D and 2D) systems~\cite{knoll2020negative,sharma2020sign,cortijo2016linear,zhou2019valley,das2021intrinsic,pal2021berry,han2022orbital}. The valley-contrasting OMM, which considerably arises from the self-rotation of Bloch wave packets~\cite{chang2008berry}, shifts the energy dispersion through a Zeeman-like coupling to the magnetic field and hence modifies the band velocity.

In this paper, we study the impact of considering the OMM on the magnetoresistance (MR) in 2D multivalley systems following an exact solution to the semiclassical Boltzmann equation. In previous studies~\cite{zhou2019valley,das2021intrinsic}, the magnetotransport properties in these systems were found solving the Boltzmann transport equation in simple relaxation-time approximation. Recently Xiao $et\, al$ ~\cite{xiao2020linear} showed that the intra-scattering effects of the semiclassical dynamics (excluded in the relaxation-time approximation) which are embodied in geometric features of the wave function can be as important as the inter-scattering effects in weak-field magnetotransport studies. These terms can augment the impact of both the Berry curvature and the OMM in collision integral. It was also proposed that a strong enough intervalley scattering can substantially affect the results owing to the valley-contrasting nature of the OMM in inversion-broken 2D systems and 3D Weyl semimetals. These proposals were also approved in other studies on the magnetotransport in 3D Weyl semimetals where an exact solution of the Boltzmann equation (involving in-scattering terms) and a sufficiently strong intervalley scattering could describe the shift of the longitudinal magnetoconductance from positive to negative~\cite{knoll2020negative,sharma2020sign}.

While previous studies on the magnetotransport in 2D multivalley systems with time-reversal symmetry considered the OMM effect in a simple relaxation-time approximation~\cite{zhou2019valley,das2021intrinsic}, it, however, remains to be understood how the in-scattering terms can modify the results in 2D. On the other hand, the impact of the intervalley scattering in the presence of the OMM on the MR of 2D multivalley systems has yet been unexplored. These remain two vital questions to be addressed in this paper.

The significant consequences of the OMM are not only limited to time-reversal-symmetric systems; it can strongly affect the magnetotransport in time-reversal-broken systems. Along with the anomalous Hall effect, which occurs in the absence of a magnetic field due to the unequal Berry curvatures in two valleys~\cite{karplus1954hall,sinitsyn2007semiclassical,nagaosa2010anomalous}, we expect a linear MR (with respect to the magnetic field) in these systems as a consequence of the Onsager's reciprocity relations. The MR in 2D time-reversal-broken multivalley systems has been studied in Ref.~\cite{sekine2018valley}, but the important contribution of the OMM was unconsidered in their calculations and a comprehensive study of the MR in the presence of the OMM in these systems is lacking. This is another important issue to be addressed here.

Our calculations show that when the Fermi level in a 2D inversion-asymmetric system lies far from the band edge, the inclusion of in-scattering terms considerably affects the MR results and can even switch the sign of the MR. While in this case, the presence of an intervalley scattering can effectively increase the absolute value of the MR, for a low-density regime it can cause a sharp drop in MR when one of the valleys is depleted in a certain magnetic field. In the case of a time-reversal-broken system, which is assumably produced by inducing valley polarization in the system, we find that the inclusion of the OMM leads to a negative intrinsic Hall conductivity in contrast to the anomalous Hall conductivity which changes sign depending on the valley polarization sign (negative for $\delta\varepsilon_{\rm F}<0$ and positive for $\delta\varepsilon_{\rm F}>0$). The MR dependence on the magnetic field also varies substantially in the presence of the OMM such that in this case, for $B>0$, the MR remains negative for $\delta\varepsilon_{\rm F}<0$, but it changes sign from positive to negative for $\delta\varepsilon_{\rm F}>0$. We show that this latter case is much more sensitive to the size of the valley polarization while the MR changes only slightly in the former case. 

The rest of the paper is organized as follows. In section II, we describe the minimal effective Hamiltonian used to describe the 2D inversion-broken multivalley system in this work. In section III, the semiclassical Boltzmann framework and the procedure to find the exact solution to that is explored. Section IV is devoted to discussing the magnetotransport results for both time-reversal-symmetric ($\delta\varepsilon_{\rm F}=0$) and asymmetric cases ($\delta\varepsilon_{\rm F}\neq0$). Finally, our results are summarized in section V.

\section{Massive Dirac model}\label{two}

A minimal low-energy Hamiltonian for a 2D multivalley system is the massive Dirac model where the gap has broken the inversion symmetry and therefore we have of two valleys connected by time-reversal symmetry. This can describe the graphene on a substrate~\cite{giovannetti2007substrate,zhou2007substrate,yankowitz2012emergence} or a monolayer transition metal dichalcogenide (TMDC), or a hybrid structure composed of a graphene layer and another 2D layer~\cite{naimer2023twist} and is given by~\cite{zhou2019valley,sekine2018valley,xiao2007valley,xiao2012coupled,faridi2020many},
\begin{equation}\label{hamiltonian}
\hat{H}=\hbar v_F(\tau k_x\hat{\sigma_x}+k_y\hat{\sigma_y})+\Delta\hat{\sigma_z},
\end{equation} 
where $\tau=\pm1$ is the valley index, $\hat{\sigma}$ denotes the Pauli matrices acting in pseudospin space, $v_F$ is the bare Fermi velocity and $2\Delta$ is the band gap. The energy dispersion of the conduction(+) and the valence(-) bands are then given by ${\varepsilon_{ k}=\pm\sqrt{(\hbar v_F )^2\vert k\vert^2+\Delta^2}}$. We consider an electron-doped system and hence the Bloch vectors of the conduction band in two valleys are given by
\begin{equation}
\vert u^{\tau}_k\rangle=\binom{\tau \cos\frac{\theta}{2}e^{-i\tau\phi}}{\sin\frac{\theta}{2}},\label{eigc}
\end{equation} 
where $\theta$ and $\phi$ are defined such that $\cos\theta=\frac{\Delta}{\varepsilon_{ k}}$, $\sin\theta\sin\phi=\frac{\hbar v_Fk_y}{\varepsilon_{ k}}$ and $\sin\theta\cos\phi=\frac{\hbar v_Fk_x}{\varepsilon_{ k}}$.

In 2D systems, the Berry curvature is along the $\hat{z}$ and for each band it is given by ${\Omega^{\tau}_{k_z}}=\partial_{k_x}A^{\tau}_{k_y}-\partial_{k_y}A^{\tau}_{k_x}$ with $A^{\tau}_{j}=\langle u^{\tau}_k\vert i\partial_{j}u^{\tau}_k\rangle$ being the $j^{th}$ component of the Berry connection $\mathbold{A^{\tau}_k}$ of the band and $\vert u^{\tau}_k\rangle$ is the Bloch vector of the same band~\cite{xiao2010berry}. For a system described by the Hamiltonian~\eqref{hamiltonian}, the Berry curvature in the conduction band is given by
\begin{equation}\label{berry}
\bm{\mathrm{\Omega}}^{\tau}_{k_z}=-\frac{\hbar^2v_F^2\Delta}{2\varepsilon_k^3}\tau{\hat{z}}.
\end{equation} 
In this paper, we also consider the OMM which can be described as the self-rotation of the Bloch wave packets in each band whose $i^{th}$ component is given by ${{m^{\tau}_i}=\varepsilon_{ijk}(-i\frac{e}{2\hbar})\langle \partial_j u^{\tau}\vert \hat{H}-\varepsilon\vert \partial_{k}u^{\tau}\rangle}$ and same as the Berry curvature for a 2D system, it is a vector along the $\hat{z}$~\cite{xiao2007valley,chang2008berry}. Using this definition, the OMM of the present system in the conduction band is given by~\cite{chang1996berry,chang2008berry, fuchs2010topological}
\begin{equation}\label{omm}
\bm m^{\tau}_{k_z}=-\frac{e\hbar v_F^2\Delta}{2\varepsilon_k^2}\tau\hat{z}.
\end{equation} 
In the presence of a perpendicular magnetic field, the OMM couples to the magnetic field and shifts the energy as $\tilde{\varepsilon}_k=\varepsilon_k-\mathbf{m}^\tau_k \cdot\bf B$. The Berry curvature and OMM provide valley-contrasting effects on transport because of their dependency on the valley $\tau$. For $B>0$, the valley-contrasting nature of the OMM leads to an upward shift in one valley ($\mathbold K$) and a downward shift in the other valley ($\mathbold{K'}$). For a low-doping small-gap system, the OMM is about 30 times the Bohr magneton~\cite{xiao2007valley} and therefore we have ignored the spin splitting in the presence of the magnetic field. Following the OMM-induced energy shift, the band velocity also changes as $\mathbold{\tilde{v}_k}=\partial_{\hbar\mathbold k}\tilde{\varepsilon}_k$.

\section{Boltzmann framework}

The semiclassical equations of motion for Bloch wave packets are~\cite{xiao2010berry,sundaram1999wave}
 \begin{eqnarray}
D_k\mathbold{\dot{r}}= \mathbold{\tilde{v}_k}+\frac{e}{\hbar}\mathbold{E}\times\bf{\Omega},\label{m1}\\
D_k\mathbold{\dot{k}}=-\frac{e}{\hbar}[\mathbold{E}+ \mathbold{\tilde{v}_k}\times\mathbold{B}],\label{m2}
\end{eqnarray} 
where $D_k=1+\frac{e}{\hbar}\mathbold{B}\cdot\bf{\Omega}$ is the Berry curvature induced modified measure for the number of states in the reciprocal space~\cite{xiao2005berry}.

The semiclassical Boltzmann equation for valley $\tau$ reads~\cite{ashcroft1976solid}
 \begin{equation}\label{Boltz}
\frac{\partial f_k^{\tau}}{\partial t}+\mathbold{\dot{r}}^{\tau}\cdot\bm{\nabla}_r f_k^{\tau}+\mathbold{\dot{k}}^{\tau}\cdot\bm{\nabla}_k f_k^{\tau}=\mathcal{I}_{col}[f_k^{\tau}].
\end{equation} 
Here $f_k^{\tau}$ is the nonequilibrium distribution function written as $f_k^{\tau}=f^{eq}(\tilde{\varepsilon}_k^{\tau})+g_k^{\tau}$ where $f^{eq}(\tilde{\varepsilon}_k^{\tau})$ is the equilibrium Fermi-Dirac distribution function and $g_k^{\tau}$ is the deviation from equilibrium. Note that in the presence of a perpendicular magnetic field $\mathbold B$ and for a system with a finite Berry curvature $\bf{\Omega}$, the equilibrium distribution function should be calculated at $\tilde{\varepsilon}_k^{\tau}$. The term on the right hand side of the equation is the collision integral. It incorporates the effect of the impurity scatterings in the system which tends to relax the nonequilibrium distribution function toward the Fermi surface and for an elastic scattering, it is written as 
 \begin{equation}\label{col}
\mathcal{I}_{col}[f_k^{\tau}]=\sum_{k',{\tau}'} W_{kk'}^{{\tau}{\tau}'}(\eta f_{k'}^{{\tau}'}-f_k^{\tau}),
\end{equation} 
where the first and second terms are expressing the in-scattering and out-scattering contributions respectively. Here, we have attached a factor ${\eta=\lbrace 0,1\rbrace}$ to the in-scattering terms in the collision integral to trace the effect of this contribution in our numerical results. Considering only the elastic scatterings, the scattering rate $W_{kk'}^{\tau{\tau}'}$
within the lowest order of the Born approximation reads 
 \begin{equation}\label{scatl}
W_{kk'}^{{\tau}{\tau}'}=\frac{2\pi}{\hbar}\frac{n_i}{S}{\vert\langle u^{\tau'}_{k'}\vert V_{kk'}^{{\tau}{\tau}'}\vert u^{\tau}_k\rangle\vert}^2\delta(\tilde{\varepsilon}_{k'}^{\tau'}-\tilde{\varepsilon}_{k}^{\tau}),
\end{equation} 
where $n_i$ is the impurity density and $S$ is the area of the 2D system. We set the impurity density such that the diluteness criteria will be satisfied for a wide range of chemical potentials.  $V_{kk'}^{{\tau}{\tau}'}$ is also the scattering matrix element for nonmagnetic point-like scattering centers and since we are considering the elastic scatterings, it is not a function of momentum and it reduces to $V^{{\tau}{\tau}'}\mathbb{I}$. Here we consider uncorrelated point scatterers, so we can consider both intravalley $V_{\mathrm{intra}}$ and intervalley $V_{\mathrm{inter}}$ scattering amplitudes~\cite{knoll2020negative,sharma2020sign,sekine2018valley}. The overlap of the Bloch vectors (Eq.~\eqref{eigc}) in the scattering rate also gives
 \begin{equation}\label{ovl}
 \begin{split}
&{\vert\langle u^{\tau'}_{k'}\vert V_{kk'}^{{\tau}{\tau}'}\vert u^{\tau}_k\rangle\vert}^2=\\
&\frac{1}{2} \vert  V^{{\tau}{\tau}'}\vert^2[1+\cos\theta\cos\theta'+\tau\tau'\sin\theta\sin\theta'\cos(\tau'\phi'-\tau\phi)],
\end{split}
\end{equation}   

Now for a uniform system in its stationary state, the Boltzmann Eq.~\eqref{Boltz} reduces to 
 \begin{equation}\label{Boltz3}
-\frac{1}{D_k^{\tau}}\frac{e}{\hbar}[\mathbold{E}+\mathbold{\tilde{v}_k^{\tau}}\times\mathbold B]\cdot\bm{\nabla}_k f_k^{\tau}=\mathcal{I}_{col}[f_k^{\tau}].
\end{equation} 

To solve the above equation we follow an exact integral equation approach~\cite{vyborny2009semiclassical,faridi2016electron,zare2017thermoelectric} and we generalize it to the case of a finite perpendicular magnetic field, $\mathbold B$. Expanding the nonequilibrium distribution function $f_k^{\tau}$ to linear order in $\mathbold{E}$, we have
 \begin{equation}\label{f0}
f_k^{\tau}=f^{eq}(\tilde{\varepsilon}_k^{\tau})+E_x\partial_{E_x}f^{\tau}+E_y\partial_{E_y}f^{\tau}.
\end{equation} 
Considering $\mathbold k=k(\cos\phi,\sin\phi)$ and $\mathbold E=E(\cos \xi,\sin \xi)$, the nonequilibrium part of the distribution function reads
 \begin{equation}\label{f}
g^{\tau}(\phi,\xi)=E[A^{\tau}(\phi)\cos\xi+B^{\tau}(\phi)\sin\xi],
\end{equation} 
where $A^{\tau}(\phi)=\partial_{E_x}g^{\tau}$ and $B^{\tau}(\phi)=\partial_{E_y}g^{\tau}$.

Using the definition of $f_k^{\tau}$ and keeping only the linear terms in $\mathbold{E}$, Eq.~\eqref{Boltz3} reduces to
 \begin{equation}\label{Boltz4}
-\frac{1}{D_k^{\tau}}\frac{e}{\hbar}[\hbar\mathbold{E}\cdot\mathbold{\tilde{v}_k^{\tau}}(\frac{\partial f^{eq}}{\partial \tilde{\varepsilon}_k^{\tau}})+\mathbold{\tilde{v}_k^{\tau}}\times\mathbold B\cdot\bm{\nabla}_kg_k^{\tau}]=\mathcal{I}_{col}[g_k^{\tau}].
\end{equation} 

For isotropic Fermi surfaces, the velocity vector $\mathbold{\tilde{v}_k^{\tau}}$ is in the $\hat{k}$ direction and therefore the second term on the left-hand side of the above equation is given by $-\frac{\tilde{v}_k^{\tau}B}{k}\frac{\partial g^{\tau}}{\partial\phi}$. Keeping this in mind and plugging  \eqref{f} into \eqref{Boltz4}, we will end up with the following equations

 \begin{equation}
  \begin{split}
&\cos\phi-F^{\tau}(k)\partial_{\phi}a^{\tau}(\phi)=G^{\tau}(k)a^{\tau}(\phi)-\\
&\sum_{{\tau}'}\int [H^{\tau\tau'}(k)+\tau\tau'I^{\tau\tau'}(k)\cos(\tau'\phi'-\tau\phi)]a^{\tau'}(\phi')d\phi',\label{B3}
\end{split}
\end{equation}
and
 \begin{equation}
\begin{split}
&\sin\phi-F^{\tau}(k)\partial_{\phi}b^{\tau}(\phi)=G^{\tau}(k)b^{\tau}(\phi)-\\
&\sum_{{\tau}'}\int [H^{\tau\tau'}(k)+\tau\tau'I^{\tau\tau'}(k)\cos(\tau'\phi'-\tau\phi)]b^{\tau'}(\phi')d\phi',\label{B4}
\end{split}
\end{equation} 
with
 \begin{align}
&F^{\tau}(k)=\frac{e\tilde{v}_k^{\tau}B}{\hbar k},\label{F}\\
&G^{\tau}(k)=\frac{S}{4\pi^2}D_k^{\tau}\sum_{\tau'}\int d^2k' D_{k'}^{\tau'} W_{kk'}^{{\tau}{\tau}'},\label{G}\\
&H^{\tau\!\tau'}\!\!(k)\!\!=\!\!\frac{n_i}{4\pi\hbar}\!D_k^{\tau}\!\!\!\int\!\! k'\!dk'\! D_{k'}^{\tau'}\frac{\tilde{v}_{k'}^{\tau'}}{\tilde{v}_k^{\tau}}\vert\! V^{{\tau}{\tau}'}\!\!\vert^2\!(1\!\!+\!\!\frac{\Delta^2}{\varepsilon(k)\varepsilon(k')}\!)\delta(\!\tilde{\varepsilon}_{k'}^{\tau'}\!\!-\!\tilde{\varepsilon}_{k}^{\tau}\!),\label{H}\\
&I^{\tau\tau'}(k)=\frac{n_i}{4\pi\hbar}D_k^{\tau}\!\!\!\int\!\! k'dk'\! D_{k'}^{\tau'}\frac{\tilde{v}_{k'}^{\tau'}}{\tilde{v}_k^{\tau}}\vert V^{{\tau}{\tau}'}\!\vert^2\frac{\hbar^2v_F^2 kk'}{\varepsilon(k)\varepsilon(k')}\delta(\!\tilde{\varepsilon}_{k'}^{\tau'}\!\!-\!\tilde{\varepsilon}_{k}^{\tau}\!),\label{I}
\end{align} 
where we have defined $a^{\tau}(\phi)\equiv{A^{\tau}(\phi)}/{e\tilde{v}_k^{\tau}(\frac{\partial f^{eq}}{\partial \tilde{\varepsilon}_k^{\tau}})}$ and $b^{\tau}(\phi)\equiv{B^{\tau}(\phi)}/{e\tilde{v}_k^{\tau}(\frac{\partial f^{eq}}{\partial \tilde{\varepsilon}_k^{\tau}})}$.
Note that we have replaced $\sum_{k'}$ with $\frac{S}{4\pi^2}\int d^2k' D_{k'}^{\tau'}$ to capture the Berry phase induced modification of the density of states. The integrals in Eqs.~(\ref{G})-(\ref{I}) can be analytically found due to the elastic scattering condition $\delta(\tilde{\varepsilon}_{k'}^{\tau'}-\tilde{\varepsilon}_{k}^{\tau})$ (See Appendix.~\ref{ap-one}). Solving Eq.~\eqref{B3} and \eqref{B4} for $a^{\tau}(\phi)$ and $b^{\tau}(\phi)$ and putting them back into Eq.~\eqref{f}, we will have the exact solution of the Boltzmann equation up to the leading order in $\mathbold{E}$.

To find $a^{\tau}(\phi)$ and $b^{\tau}(\phi)$, we write them in the form of Fourier series
 \begin{eqnarray}\label{a}
 &&
 \begin{split}
a^{\tau}(\phi)&=a_0^{\tau}+\sum_{n=1}\{a_{cn}^{\tau}\cos(n\phi)+a_{sn}^{\tau}\sin(n\phi)\}
\end{split}
\\[10pt]
&&
\begin{split}
b^{\tau}(\phi)&=b_0^{\tau}+\sum_{n=1}\{b_{cn}^{\tau}\cos(n\phi)+b_{sn}^{\tau}\sin(n\phi)\}
\end{split}
\end{eqnarray} 

Plugging $a^{\tau}(\phi)$ and $b^{\tau}(\phi)$ in Eqs.~\eqref{B3} and \eqref{B4} and after a straightforward algebra, we will end up with a system of coupled equations (the details can be found in Appendix.~\ref{ap-two} ) and finally the field induced correction to the distribution function reads
 \begin{equation}\label{g}
 \begin{split}
&g^{\tau}=e\tilde{v}_k^{\tau}(\frac{\partial f^{eq}}{\partial \tilde{\varepsilon}_k^{\tau}})E\times\\
&\bigl\lbrace [a_{c1}^{\tau}\cos\phi+a_{s1}^{\tau}\sin\phi]\cos\xi+ [b_{c1}^{\tau}\cos\phi+b_{s1}^{\tau}\sin\phi]\sin\xi\bigr\rbrace.
\end{split}
\end{equation} 
which is the main quantity for evaluating the linear response conductivities.
\section{Magnetotransport results}

Having found the distribution function of the system, $f^{eq}(\tilde{\varepsilon}_k^{\tau})+g_k^{\tau}$, we can now calculate the longitudinal and Hall conductivities. For the current density we have
\begin{equation}\label{j0}
\mathbold j=-\frac{e}{S}\sum_{k,\tau}\mathbold{\dot{r}}^{\tau}f_k^{\tau}=-\frac{e}{4\pi^2}\sum_{\tau}\int d^2k D_k \mathbold{\dot{r}}^{\tau}[f^{eq}(\tilde{\varepsilon}_k^{\tau})+g_k^{\tau}].
\end{equation}
Using the definition of $\mathbold{\dot{r}}^{\tau}$ from Eq.~\eqref{m1}, we have
\begin{equation}\label{j}
\begin{split}
\mathbold j=\sum_{\tau}&\bigl[-\frac{e}{4\pi^2}\int d^2k\, \mathbold{\tilde v}_k ^{\tau}\, g_k^{\tau}\\
&-\frac{e^2}{\hbar}\frac{1}{4\pi^2}\int d^2kf^{eq}(\tilde{\varepsilon}_k^{\tau})[\mathbold{E}\times\bm{\mathrm{\Omega}}^{\tau}_{k_z}]\bigr]+\it{O(E^2)}.
\end{split}
\end{equation}
Now plugging~\eqref{g} in Eq.~\eqref{j}, the longitudinal conductivity ($\xi=0$) is given by
\begin{equation}\label{xx}
\begin{split}
\sigma_{xx}\!&=\!-\frac{e^2}{4\pi^2}\!\sum_{\tau}\!\!\int\!\! d^2k ({\tilde v}_k ^{\tau})^2(\frac{\partial f^{eq}}{\partial \tilde{\varepsilon}_k^{\tau}})\!\cos\phi [a_{c1}^{\tau}\cos\phi+a_{s1}^{\tau}\sin\phi]\\
&=-\frac{e^2}{\hbar}\frac{1}{4\pi}\sum_{\tau}\tilde{k}^{\tau}{\tilde v}_{\tilde{k}^{\tau}} ^{\tau}a_{c1}^{\tau},
\end{split}
\end{equation}
and for the Hall conductivity ($\xi=\pi/2$) we have
\begin{equation}\label{xy}
\begin{split}
\sigma_{xy}\!&=\!-\frac{e^2}{4\pi^2}\!\sum_{\tau}\!\!\int\!\! d^2k({\tilde v}_k ^{\tau})^2(\frac{\partial f^{eq}}{\partial \tilde{\varepsilon}_k^{\tau}})\!\cos\phi [b_{c1}^{\tau}\!\cos\phi+\!b_{s1}^{\tau}\!\sin\phi]\\
&-\frac{e^2}{\hbar}\frac{1}{4\pi^2}\sum_{\tau}\int d^2kf^{eq}(\tilde{\varepsilon}_k^{\tau})\Omega^{\tau}_{k_z}\\
&=-\frac{e^2}{\hbar}\frac{1}{4\pi}\sum_{\tau}\lbrace\tilde{k}^{\tau}{\tilde v}_{\tilde{k}^{\tau}} ^{\tau}b_{c1}^{\tau}-\tau[\frac{\Delta}{\varepsilon_{\tilde{k}^{\tau}}}-1]\rbrace=\sigma_{xy}^{(i)}+\sigma_{xy}^{(ii)},
\end{split}
\end{equation}
where $\tilde{k}^{\tau}$ is the modified Fermi wave vector found from the condition $\varepsilon_k-m_{k_z}^{\tau}B=\varepsilon_{\rm F}$. In the absence of the OMM, Eq.~\eqref{xx} and the first term in Eq.~\eqref{xy}, $\sigma_{xy}^{(i)}$, are the ordinary longitudinal and Hall responses induced by Lorentz force and the second term in Eq.~\eqref{xy}, $\sigma_{xy}^{(ii)}$, is the quantum mechanical intrinsic Hall conductivity. Owing to the fact that the Berry curvature in $\mathbold K$ and $\mathbold{K'}$ valleys is equal and opposite in sign, this term is zero for time-reversal symmetric systems. In the case of time-reversal broken systems, this term gives the intrinsic anomalous Hall conductivity due to unequal Berry curvatures in two valleys. Now, when the OMM effect is considered, not only the ordinary and Hall conductivities are modified, but also the opposite shifting of the bands in valleys and subsequent unequal Berry curvatures (see Fig. \ref{fig1}(a)) lead to a $B$-dependent finite value for the intrinsic term in both time-reversal symmetric and asymmetric cases. 

Next, we use these conductivities to find the MR in the system defined as
\begin{equation}\label{mr}
\mathrm{MR}=\frac{\rho_{xx}(B)-\rho_{xx}(0)}{\rho_{xx}(0)},
\end{equation}
where the longitudinal resistivity is given by
\begin{equation}\label{ro}
\rho_{xx}=\frac{\sigma_{xx}}{\sigma_{xx}^2+\sigma_{xy}^2}.
\end{equation}
Below, we calculate the magnetoconductance and MR for the massive Dirac model described in Sec.~\ref{two} and discuss the results.

\subsection {Initially valley-degenerate case: $\mathbold{\delta\varepsilon_{\rm F}=0}$}

The effect of the OMM on the magnetotransport properties in 2D multivalley systems with similar population of valleys has been discussed before applying the simple relaxation-time approximation and ignoring the in-scattering terms and the possible intervalley scatterings~\cite{zhou2019valley,das2021intrinsic}. Therefore, in this subsection, we focus on the significant impact of these terms in the final MR of the system. The intervalley scattering in 2D systems is mainly controlled by sharp impurities or short-ranged scattering centers whose concentration depends on fabrication techniques. Therefore, depending on the experimental conditions and the fabrication techniques, the intervalley scattering can change substantially and get different values~\cite{chen2009defect,wu2013vapor}. In what follows, we have assumed the intravalley scattering is the same for both valleys, ${V_{\mathrm{intra}}=V_{\mathrm{11}}=V_{\mathrm{22}}}$ and to respect the valley symmetry for the intervalley scattering we have ${V_{\mathrm{inter}}=V_{\mathrm{12}}=V_{\mathrm{21}}}$. On the other hand, as we mentioned earlier, to determine the effect of the in-scattering contribution in final results, we have attached a factor ${\eta=\lbrace 0,1\rbrace}$ to the in-scattering terms in the collision integral Eq.~\eqref{col} which allows us to trace the impact of each term in the collision integral by easily switching on and off the in-scattering effect.

\begin{figure}[!ht]
 \captionsetup[subfigure]{labelformat=empty}
\centering
  \subfloat[]{\includegraphics[width=.5\linewidth]{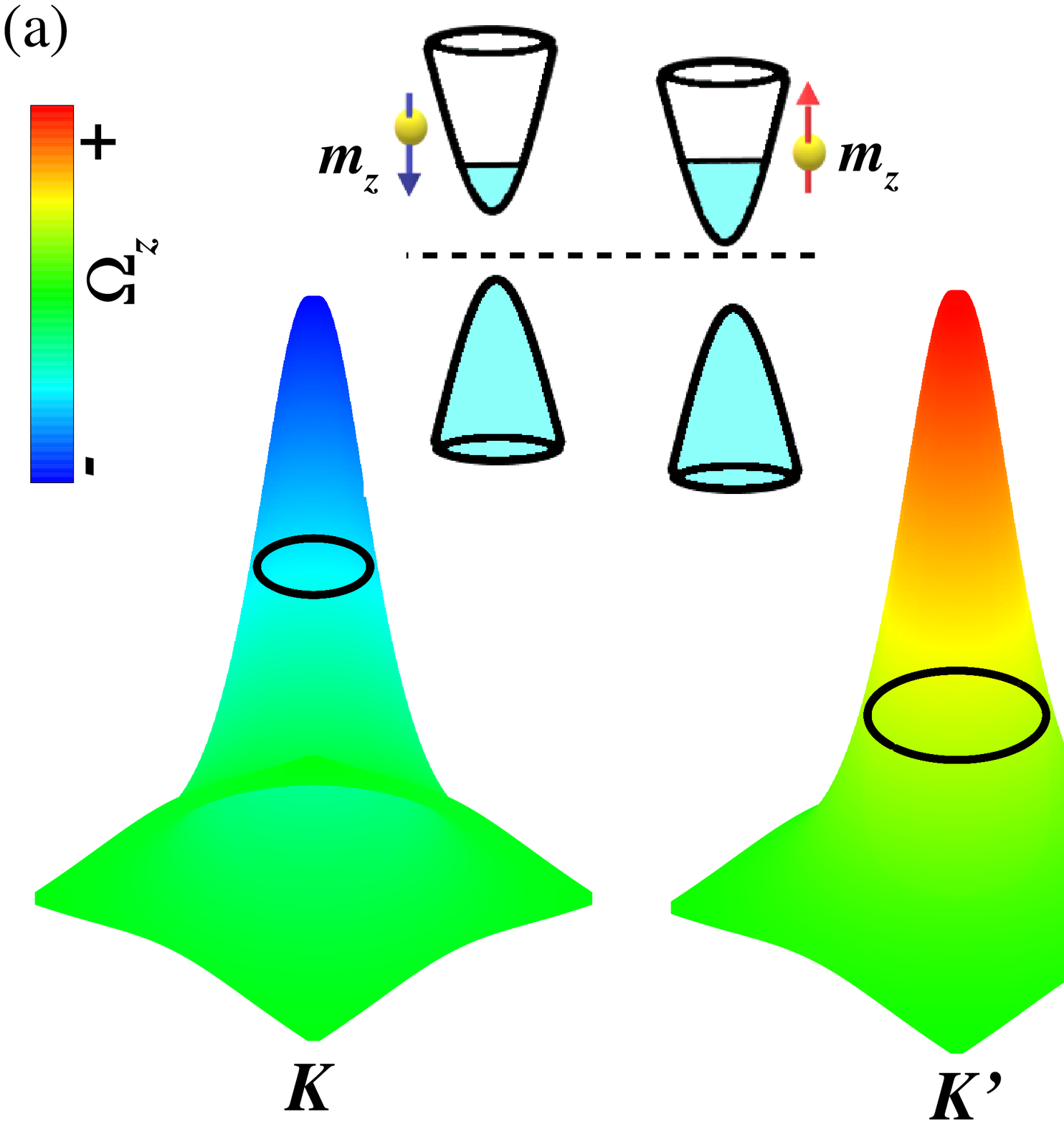}}\negthickspace\negthickspace\negthickspace
  \hfil
    \subfloat[]{\includegraphics[width=0.5\linewidth]{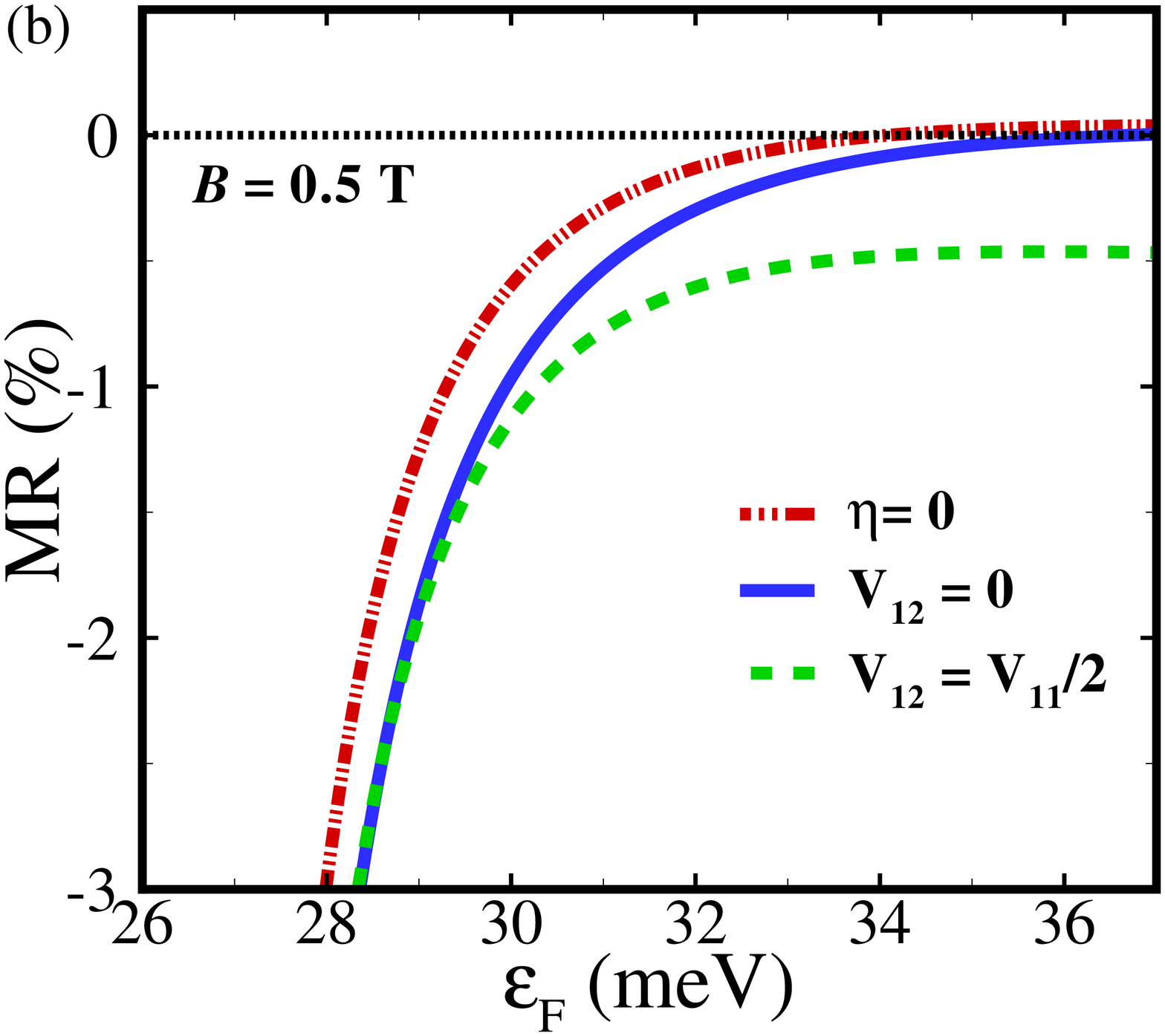}}
\par
  
    \subfloat[]{\includegraphics[width=0.5\linewidth]{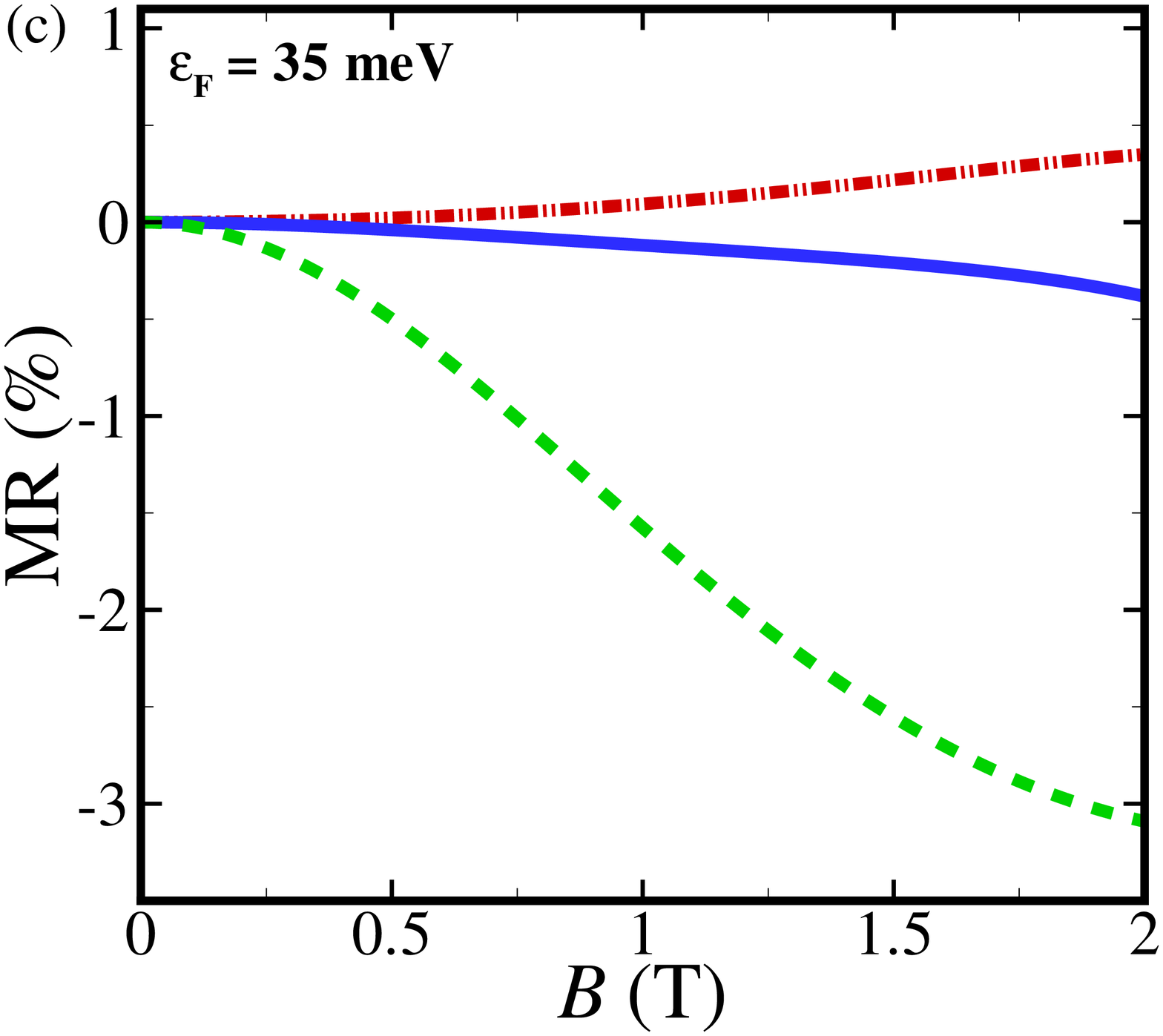}}
\hfil
    \subfloat[]{\includegraphics[width=0.5\linewidth]{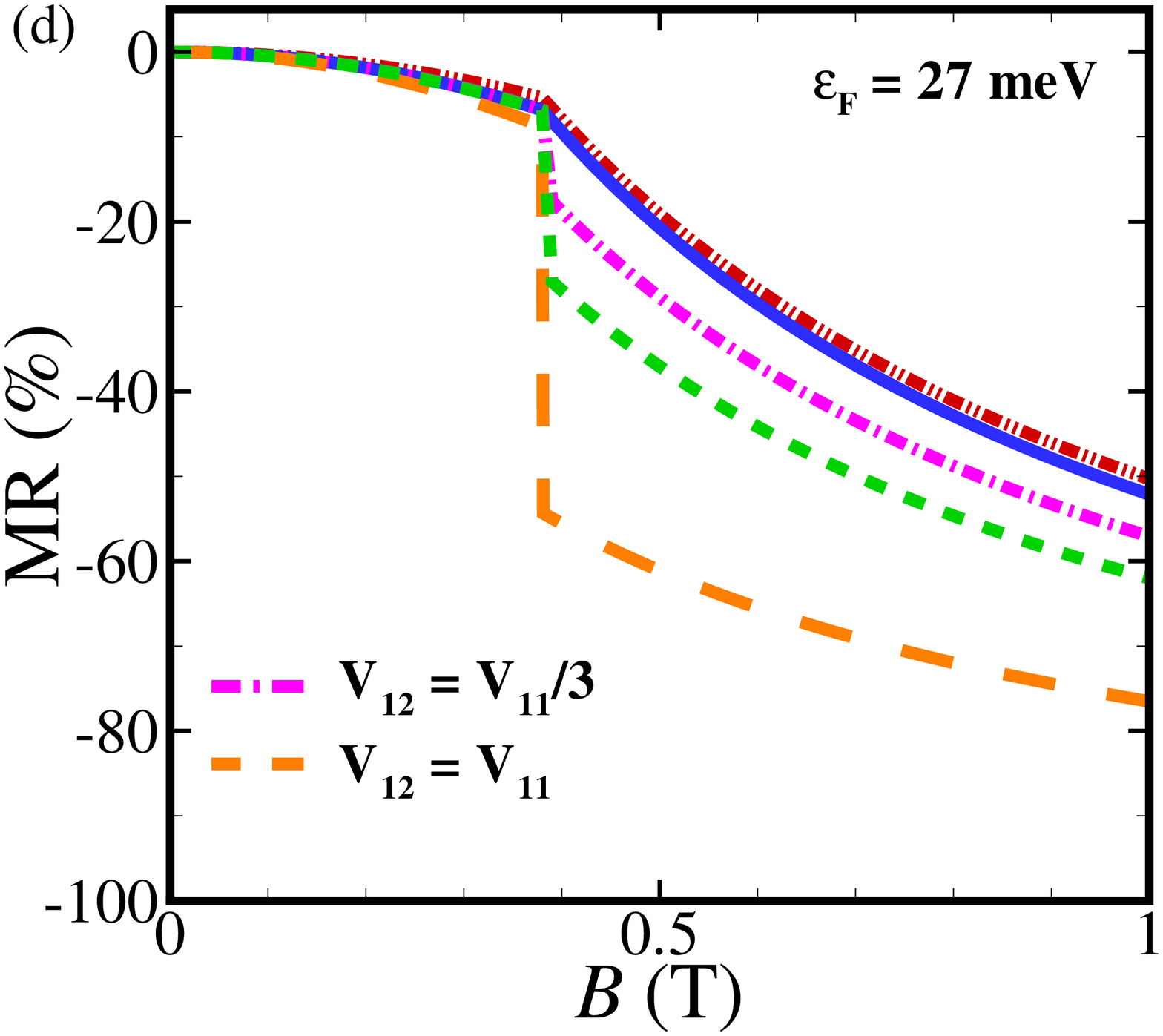}}

\caption{(Color online) (a) The opposite Berry curvatures in two valleys. The black circles indicate the Fermi surfaces which are different in two valleys due to the shift of the energy bands caused by the OMM. Here, the Fermi surfaces are plotted for $\varepsilon_{\rm F}=35$ meV and $B=2$ T. The top inset is a schematic illustration of the shifted energy bands around $\mathbold K$ and $\mathbold {K'}$ valleys and the corresponding alignment of the OMMs, $\mathbold m_z$. (b) The MR as a function of the Fermi energy, $\varepsilon_{\rm F}$ at $B=0.5$ T in the absence of in-scattering and intervalley scattering, $\eta=0$, (the red dashed-dotted curve), when in-scattering terms are considered in the collision integral but the intervalley scattering is absent (the blue solid curve), and finally in the presence of both in-scattering terms and an intervalley scattering half as large as the intravalley scattering. (c) The MR versus the magnetic field $B$ for $\varepsilon_{\rm F}= 35$ meV and (d) for $\varepsilon_{\rm F}= 27$ meV. For all the plots other than red, $\eta=1$. The parameters are set as $\hbar v_F=3\,\rm{eV\AA}$~\cite{zhou2019valley,sekine2018valley}, ${\Delta=26\,\rm{meV}}$~\cite{zhou2007substrate,yankowitz2012emergence}. Here, $n_i\vert V_{11}\vert^2$ is chosen such that for $B=0$, we have $\tau_{0}=10^{-13}\,s$~\cite{zhou2019valley,sekine2018valley}.\label{fig1}    }
 \end{figure}

To commence, we show the $\varepsilon_{\rm F}$-dependence of the MR in Fig. \ref{fig1}(b). We can see that the in-scattering term decreases the MR and has a negative contribution in the MR and at certain Fermi energies it can even change the sign of the MR from positive to negative. On the other hand, if the intervalley scattering exists in a system, its negative contribution can considerably change the MR especially for larger Fermi energies.

This can be better perceived in Fig. \ref{fig1}(c) where we illustrate the MR versus magnetic field, $B$, for $\varepsilon_{\rm F}=35$ meV. While the in-scattering term has changed the MR from positive to negative, an intervalley scattering as large as half the intravalley scattering can change the MR from $-0.4\%$ to about $-3.1\%$ at $B=2\,$T. Note that the results of the MR without in-scattering are different in our paper than those of Ref.~\cite{zhou2019valley}. This is because in previous works where the relaxation-time approximation was employed, a constant relaxation time of $\tau_{0}=10^{-13}s$ was assumed. Even in the absence of in-scattering terms, this is not accurate when we possess the magnetic field. In this case, the density of states correction due to the Berry curvature should be considered in the definition of $\tau_{0}$ such that $\tau_{0}^{-1}={\frac{S}{4\pi^2}\sum_{\tau'}\int d^2k' D_{k'}^{\tau'} W_{kk'}^{{\tau}{\tau}'}}$ ~\cite{knoll2020negative}. We can see that $\tau_{0}$ depends on the magnetic field through $D_k^{\tau}=1+\frac{e}{\hbar}\mathbold{B}\cdot\bm{\mathrm{\Omega}}^{\tau}_{k_z}$ and is not the same for two valleys. If we set $ D_{k'}^{\tau'}=1$ in the definition of $\tau_{0}$, then the results of Ref.~\cite{zhou2019valley} are reproduced.

In the case of low carrier density where the Fermi energy is near the band edge, it has been shown that when $\varepsilon_{\rm F}$ is kept constant, in a certain magnetic field one of the valleys will be depleted and a significant decrease in the MR is obtained~\cite{zhou2019valley}. Here we have found the same result and also a sharp drop at the point of depletion for a finite intervalley scattering (Fig. \ref{fig1}(d)). The size of this drop increases for more significant intervalley scatterings and it can be thought of as a measure for the strength of the intervalley scattering in the system. In fact when one of the valleys is depleted, the intervalley scattering channel is also canceled which results in a sudden decrease of the resistivity at the depletion point and hereafter the system behaves as if it doesn't experience intervalley scattering at all. But, since the zero-field resistivity $\rho_0$ is larger when $V_{12}\neq0$, the MR is considerably larger in this case.
 \begin{figure}[!ht]
 \captionsetup[subfigure]{labelformat=empty}
\centering

    \subfloat[]{\includegraphics[width=0.5\linewidth]{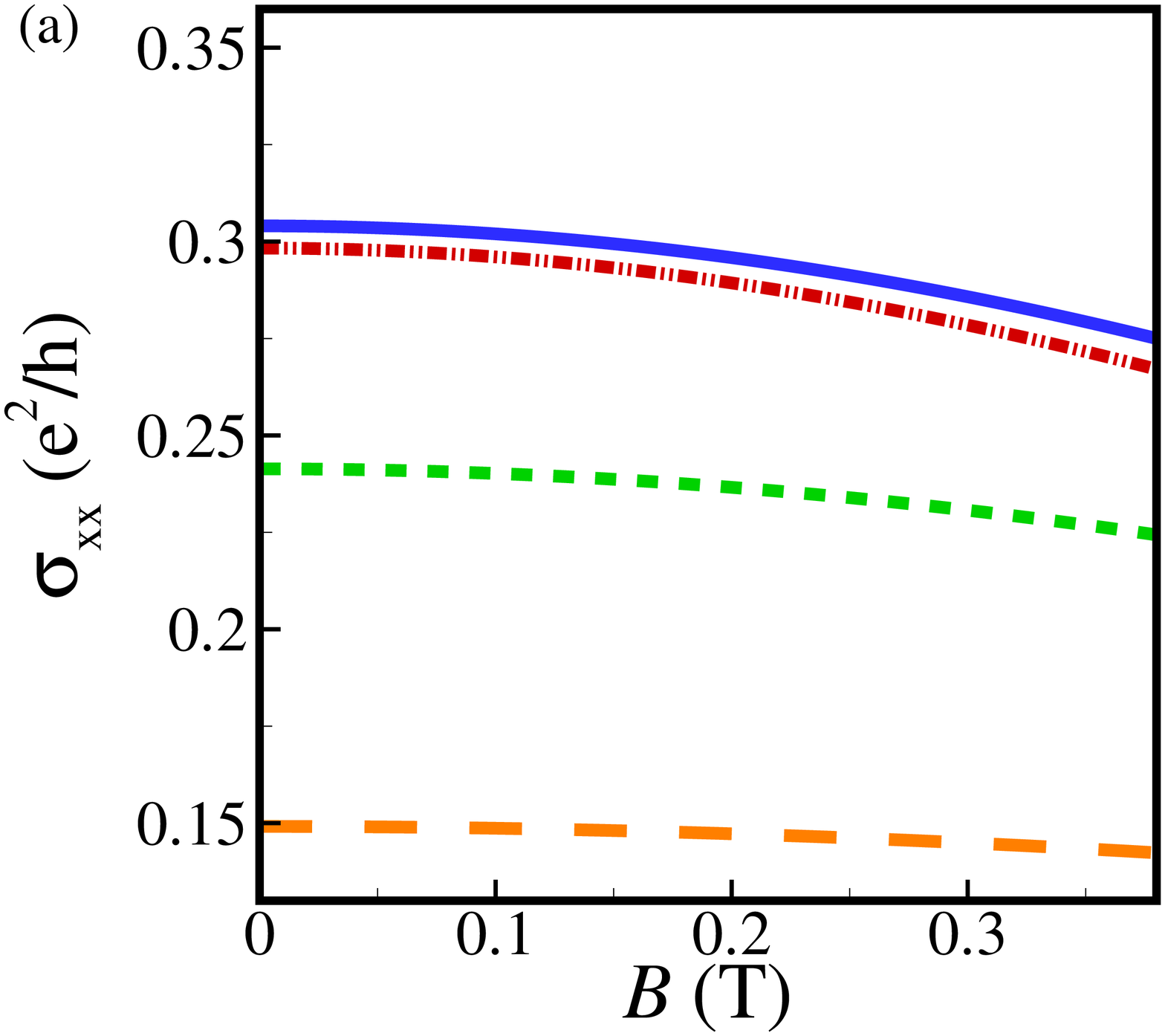}}
\hfil
    \subfloat[]{\includegraphics[width=0.5\linewidth]{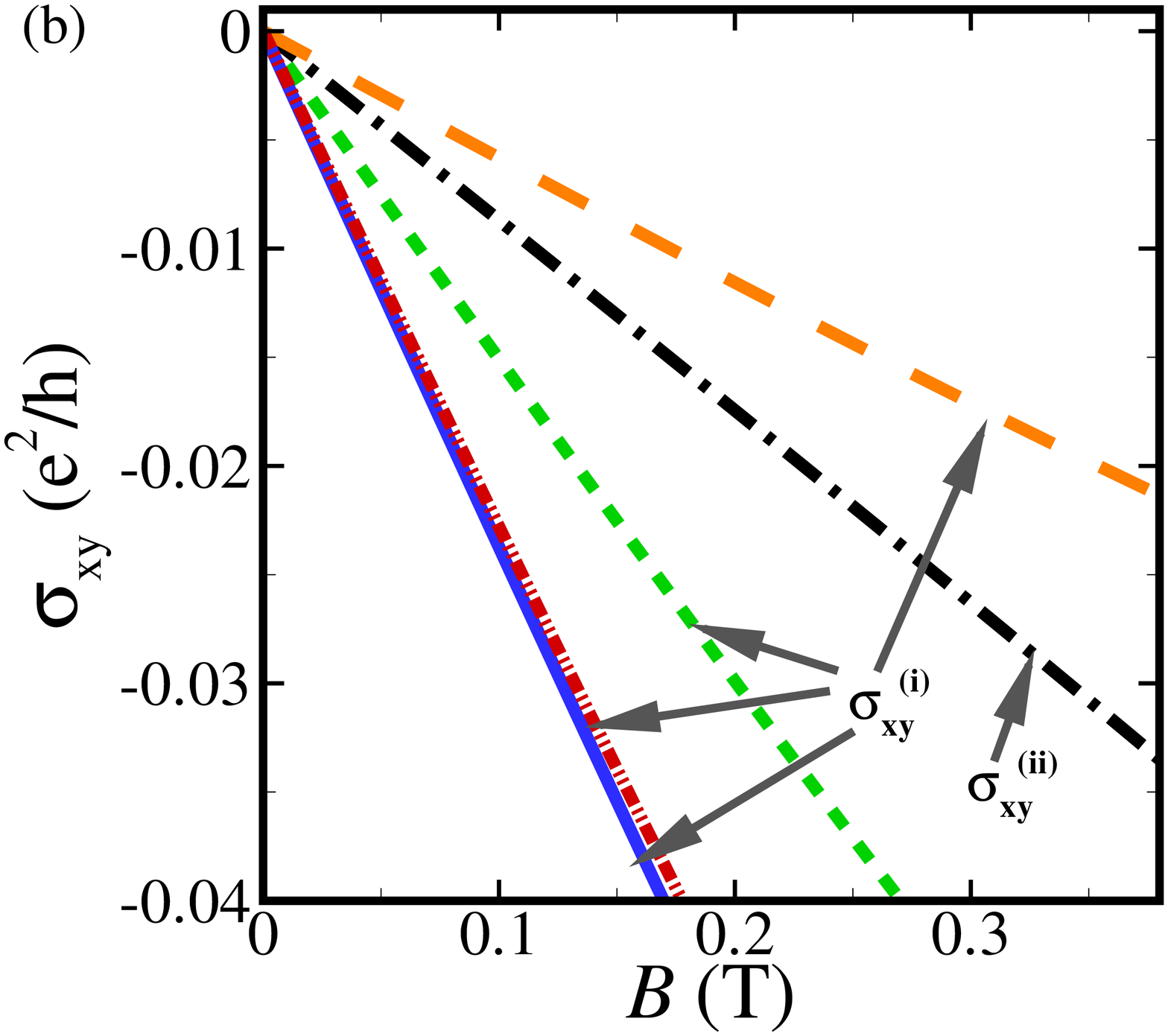}}

\caption{(Color online) (a) The longitudinal conductivity, $\sigma_{xx}$, as a function of the magnetic field, $B$, for  $\varepsilon_{\rm F}=27\,\rm{meV}$ for different values of intervalley scatterings, $V_{12}$. The colors and the format of lines are the same as Fig. \ref{fig1}(d). (b)  The Lorentz and intrinsic parts of the Hall conductivity, $\sigma_{xy}$, as a function of magnetic field, $B$ for  $\varepsilon_{\rm F}=27\,\rm{meV}$. The black dashed-dotted line is the intrinsic part of the Hall conductivity, $\sigma_{xy}^{(ii)}$, while all other lines refer to the Lorentz part of the Hall conductivity, $\sigma_{xy}^{(i)}$ with the colors and line formats similar to (a). }
    \label{fig2}
    \end{figure}
 
It is also instructive to have a quick look at the magnetoconductance in the presence of the intervalley scattering. In Fig. \ref{fig2}(a), we depict the longitudinal conductivity, $\sigma_{xx}$ as a function of magnetic field $B$ for a low-density case $\varepsilon_{\rm F}=27\,\rm{meV}$ and for reduced magnetic fields. As expected, when a new scattering channel is introduced to the system, the longitudinal conductivity decreases and a stronger intervalley scattering results in much lower longitudinal conductivity. The situation is more interesting for the Hall conductivity, $\sigma_{xy}$. In Fig. \ref{fig2}(b), we illustrate the Lorentz part of the Hall conductivity, $\sigma_{xy}^{(i)}$ for the same values of the intervalley scattering as in Fig. \ref{fig2}(a) as well as the intrinsic part $\sigma_{xy}^{(ii)}$ which is unaffected by scatterings. We can see that for $V_{12}=0$, although the Lorentz part is larger than the intrinsic part, but their magnitudes are completely comparable. For more considerable values of the intervalley scattering, the absolute value of the Lorentz part decreases highlighting the role of the intrinsic part such that for $V_{12}=V_{11}$, the largest part of the Hall conductivity belongs to the intrinsic part. We should note that since this OMM-induced intrinsic part is largest near the band edge~\cite{das2021intrinsic}, the dominant part of the Hall conductivity remains always the Lorentz term for Fermi energies far from the band bottom.

\subsection {Initially valley polarized case: $\mathbold{\delta\varepsilon_{\rm F}\neq0}$}

\begin{figure}[!ht]
 \captionsetup[subfigure]{labelformat=empty}
\centering

\subfloat[]{\includegraphics[width=0.5\linewidth]{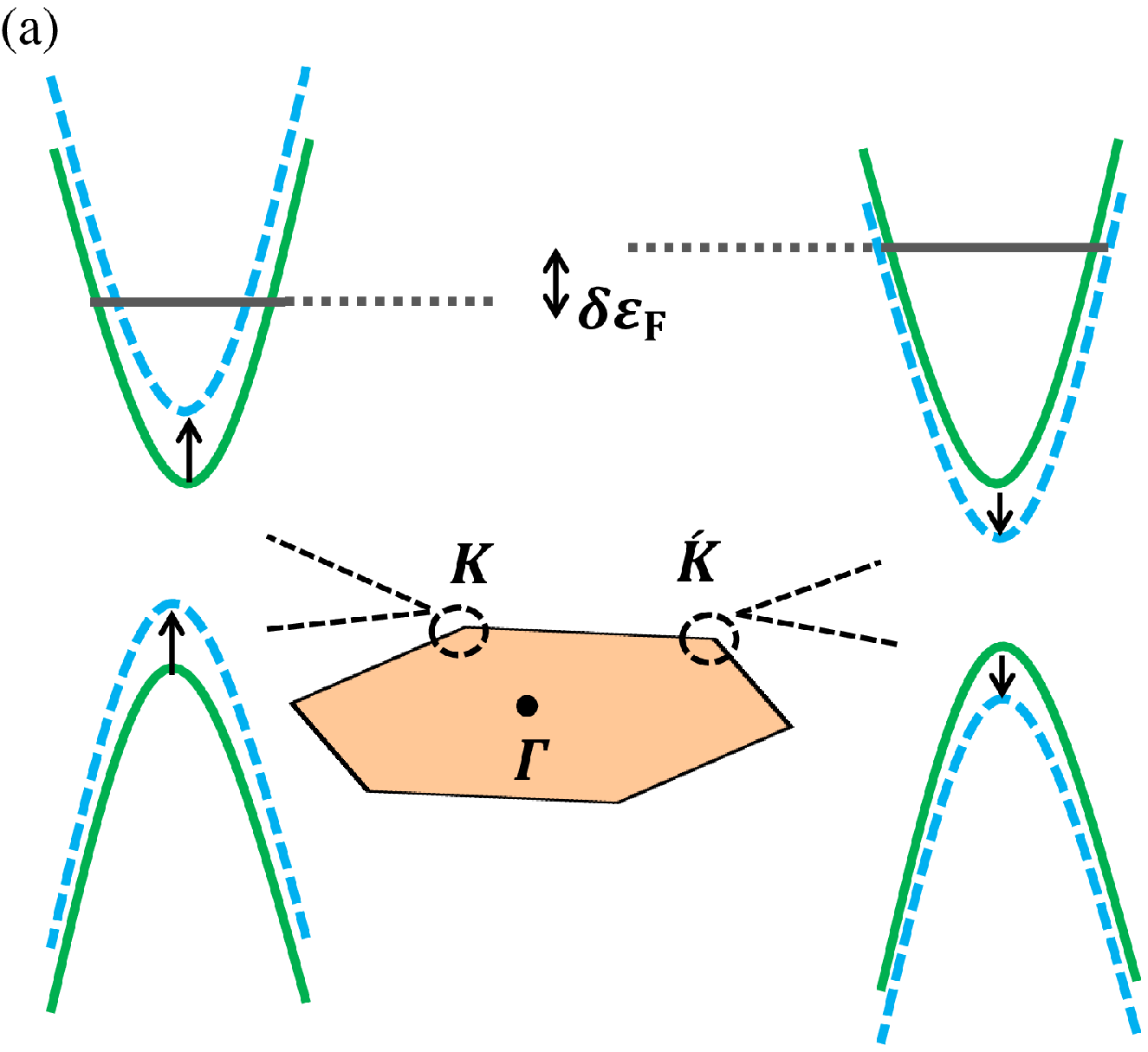}}
\par

    \subfloat[]{\includegraphics[width=0.5\linewidth]{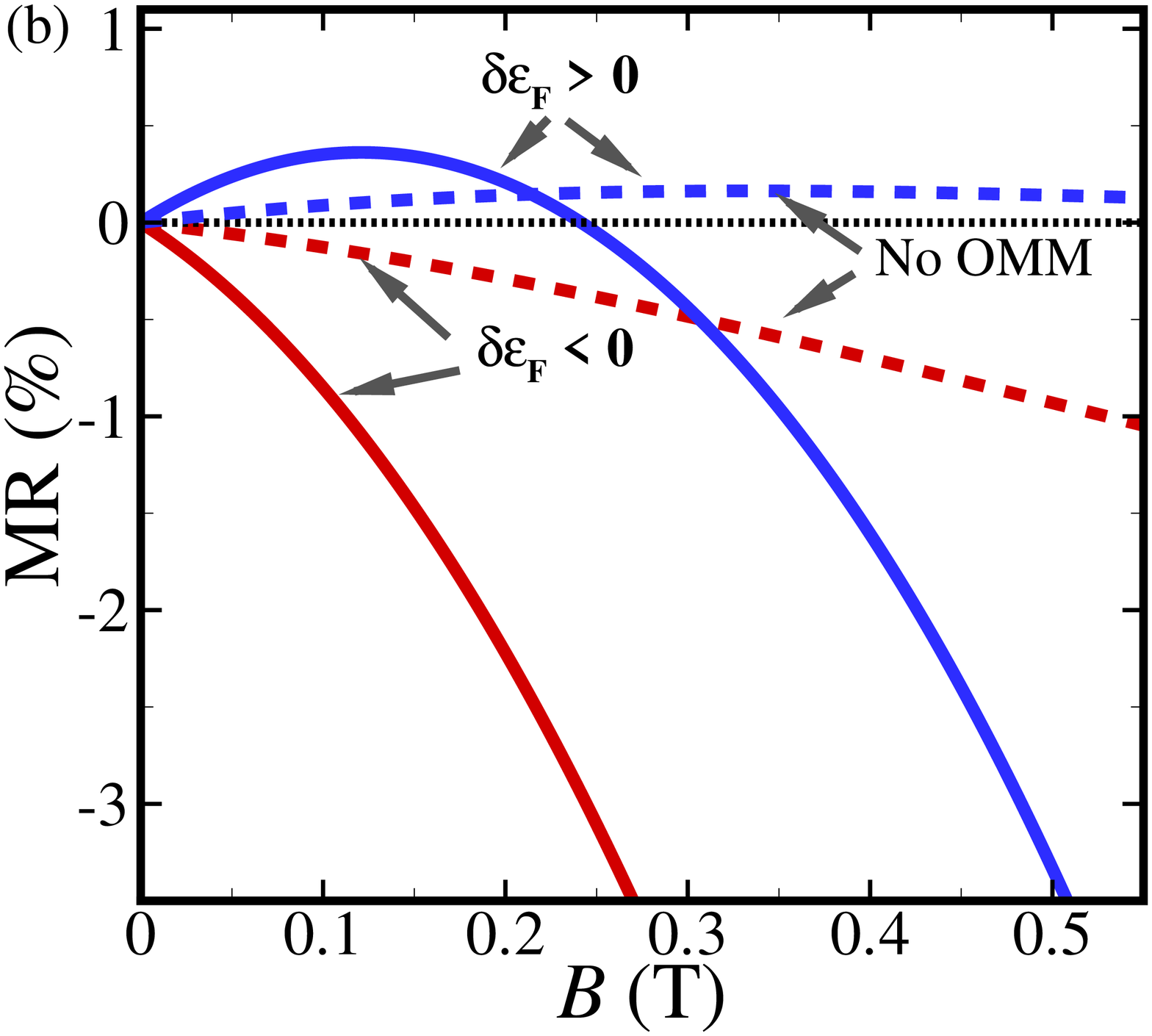}}
\hfil
    \subfloat[]{\includegraphics[width=0.5\linewidth]{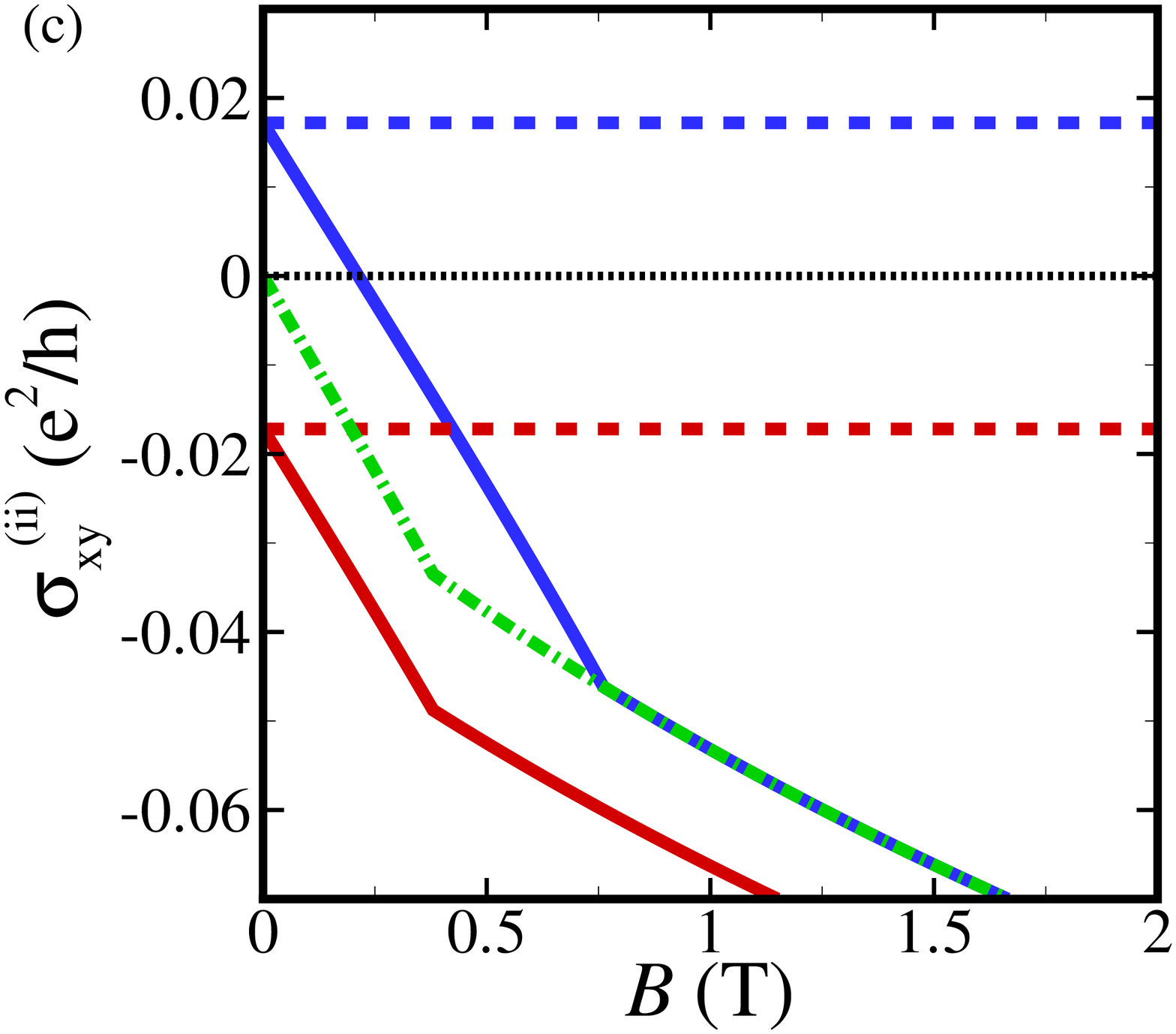}}
  
\caption{(Color online) (a) Schematic illustration of band energies in  $\mathbold{K}$ and $\mathbold{K'}$ valleys, the Fermi energy difference which results in valley polarization (negative in this case) and also the OMM-induced shift in energy bands. Here the solid lines represent the band energies in absence of magnetic field, the dashed lines correspond to the OMM-induced shifted bands, the arrows indicate the direction of the band shift, and the horizontal gray lines represent the Fermi energy in each valley.  (b) The MR versus magnetic field in the presence (solid curves) and absence (dashed curves) of the OMM. The blue curves illustrate the positive polarization case $\delta\varepsilon_{\rm F}>0$ and the red curves show the $\delta\varepsilon_{\rm F}<0$ case. (c) The intrinsic part of the Hall conductivity, $\sigma_{xy}^{(ii)}$ as a function of magnetic field, $B$. The colors and the format of lines are the same as (a) and the green dashed-dotted line shows the $\delta\varepsilon_{\rm F}=0$ case. In these panels we have assumed $V_{12}=0$ and $\delta\varepsilon_{\rm F}=1\,\rm{meV}$ such that for a positive (negative) polarization, we set $\varepsilon_{{\rm F}_{1(2)}}=28\,\rm{meV}$ and $\varepsilon_{{\rm F}_{2(1)}}=27\,\rm{meV}$. }
    \label{fig3}
    \end{figure}

The time-reversal symmetry can be broken by imposing a valley polarization on the system generating an excess of carriers in one of the valleys; for example in TMDCs this can be carried out by using circularly polarized light \cite{xiao2012coupled,zeng2012valley,mak2012control,cao2012valley}. The valley-dependent MR in this time-reversal-broken system has also been studied in Ref.~\cite{sekine2018valley}, but the important role of the OMM has been completely overlooked. Therefore, we discuss our results in this subsection with an emphasis on the OMM impact on the MR of a time-reversal-broken 2D system. In Fig. \ref{fig3}(a), we have schematically illustrated the band energies in $\mathbold{K}$ and $\mathbold{K'}$ valleys, the Fermi energy difference which results in valley polarization and also the OMM-induced shift in energy bands. Here we have shown the negative valley polarization case $\delta\varepsilon_{\rm F}<0$ ($\delta\varepsilon_{\rm F}=\varepsilon_{{\rm F}_1}-\varepsilon_{{\rm F}_2}=\varepsilon_{{\rm F}}(\mathbold K)-\varepsilon_{{\rm F}}(\mathbold {K'})$).  In Fig. \ref{fig3}(b), we can see how the inclusion of the OMM can change the MR in this case. Note that the linear-in-$B$ dependence of the MR in the weak-field regime, which is allowed in spontaneous time-reversal broken systems, can be seen in all the plots \cite{onsager1931reciprocal}.

The inclusion of the OMM can abruptly change the MR resulting in much larger values. Another interesting point is that depending on which valley is populated with the excess charge, the magnetic-field dependence of the system is very different. As we can see, for $B>0$, when OMM is considered in calculations, the MR stays negative for $\delta\varepsilon_{\rm F}<0$, while it changes sign from positive to negative values when the excess population is produced in the $\mathbold K$-valley or $\delta\varepsilon_{\rm F}>0$. As the time-reversal symmetry is broken here, the situation is reversed when we consider the $B<0$ case; which means that the sign change in MR occurs for $\delta\varepsilon_{\rm F}<0$ in this case. In fact, the nonmonotonic behavior of MR is expected whenever the sign of $\delta\varepsilon_{\rm F}$ is the same as $B$. Hereafter we consider the $B>0$ case and therefore the sign change is seen when $\delta\varepsilon_{\rm F}>0$. 

The inclusion of the OMM not only changes the energy bands (an upward shift in the $\mathbold{K}$-valley and a downward shift in the $\mathbold{K'}$-valley for $B>0$) and through that changes the dissipative currents, but it also introduces an intrinsic dissipationless Hall conductivity to the system: the second term in Eq. (\ref{xy}). In Fig. \ref{fig3}(c) we show the intrinsic contribution of the Hall conductivity, $\sigma_{xy}^{(ii)}$ as a function of the magnetic field, $B$. Note that although the OMM-induced energy shift is small compared to the Fermi energy, but since this shift is opposite in two valleys it can affect the MR considerably. This is mainly due to this intrinsic Hall conductivity which is maximized near the band edges~\cite{das2021intrinsic} and therefore for a low-doping system (same as in Fig. \ref{fig3}) it has a large value completely comparable to the ordinary Hall term and through changing the total Hall conductivity, it can strongly affect the MR even in low magnetic field. 

When the OMM is absent, the only intrinsic Hall term belongs to the anomalous Hall conductivity which is a positive value for $\delta\varepsilon_{\rm F}>0$ and a negative value for $\delta\varepsilon_{\rm F}<0$ and it is illustrated by the $B$-independent dotted lines. This term is induced in time-reversal-asymmetric systems by the opposite and unequal values of the Berry curvature in the $\mathbold K$ and $\mathbold {K'}$ valleys. When the effect of the OMM is considered, another intrinsic Hall contribution shows up. In contrast to the anomalous Hall conductivity, this OMM-induced Hall conductivity is $B$-dependent and it is always negative in accordance with the sign of the ordinary Hall conductivity induced by the Lorentz force. This is also the case for $\delta\varepsilon_{\rm F}=0$~\cite{das2021intrinsic}. In fact, the sign change seen in the MR when $\delta\varepsilon_{\rm F}>0$ is principally the result of the positive anomalous Hall conductivity at $B=0$.  When the magnetic field is turned on, the negative Hall conductivity in finite $B$ tends to decrease this positive value toward zero and after that the absolute value of the Hall conductivity increases continuously (Fig. \ref{fig3}(c)). As the Lorentz part of the Hall conductivity is also negative, this sign change in MR is expected even in the absence of OMM, but due to the slow increase of the Hall conductivity in this case, the sign change occurs in larger values of the magnetic field. In fact the negative intrinsic Hall conductivity induced by the OMM, makes the sign change happen in much smaller magnetic fields.
 \begin{figure}[!ht]
 \captionsetup[subfigure]{labelformat=empty}
\centering

    \subfloat[]{\includegraphics[width=0.5\linewidth]{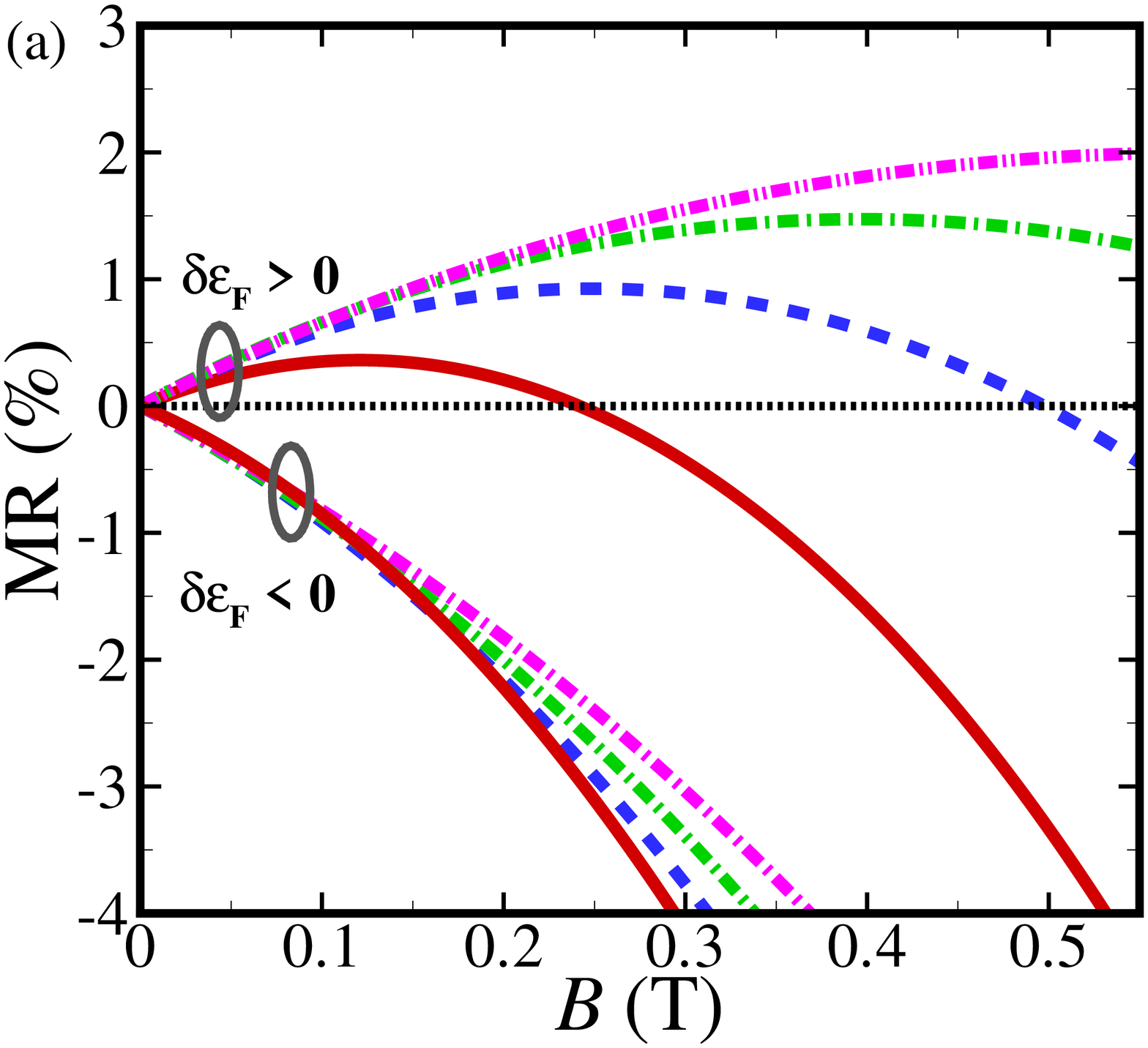}}
\hfil
    \subfloat[]{\includegraphics[width=0.5\linewidth]{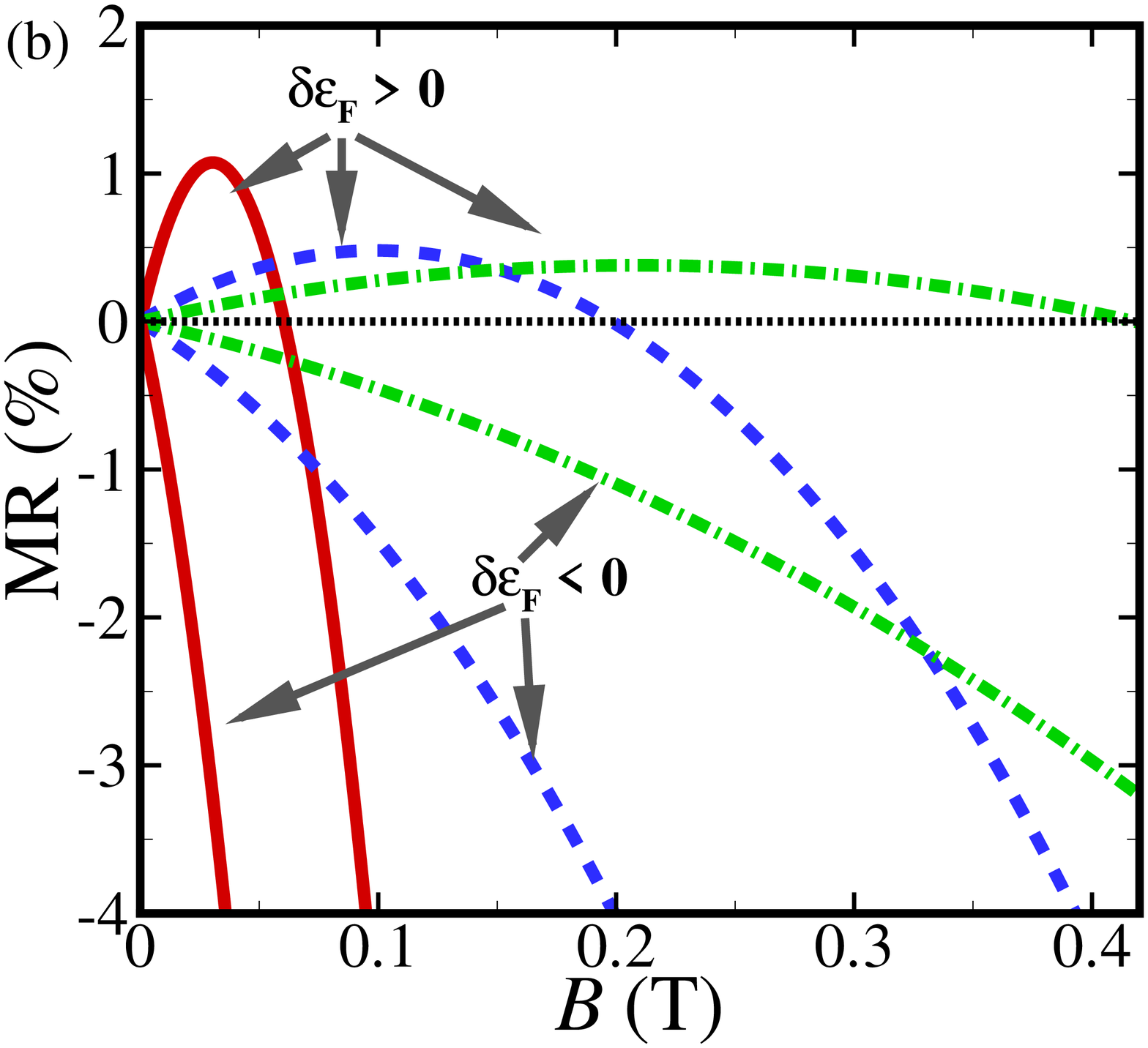}}
    
    \par
\subfloat[]{\includegraphics[width=0.5\linewidth]{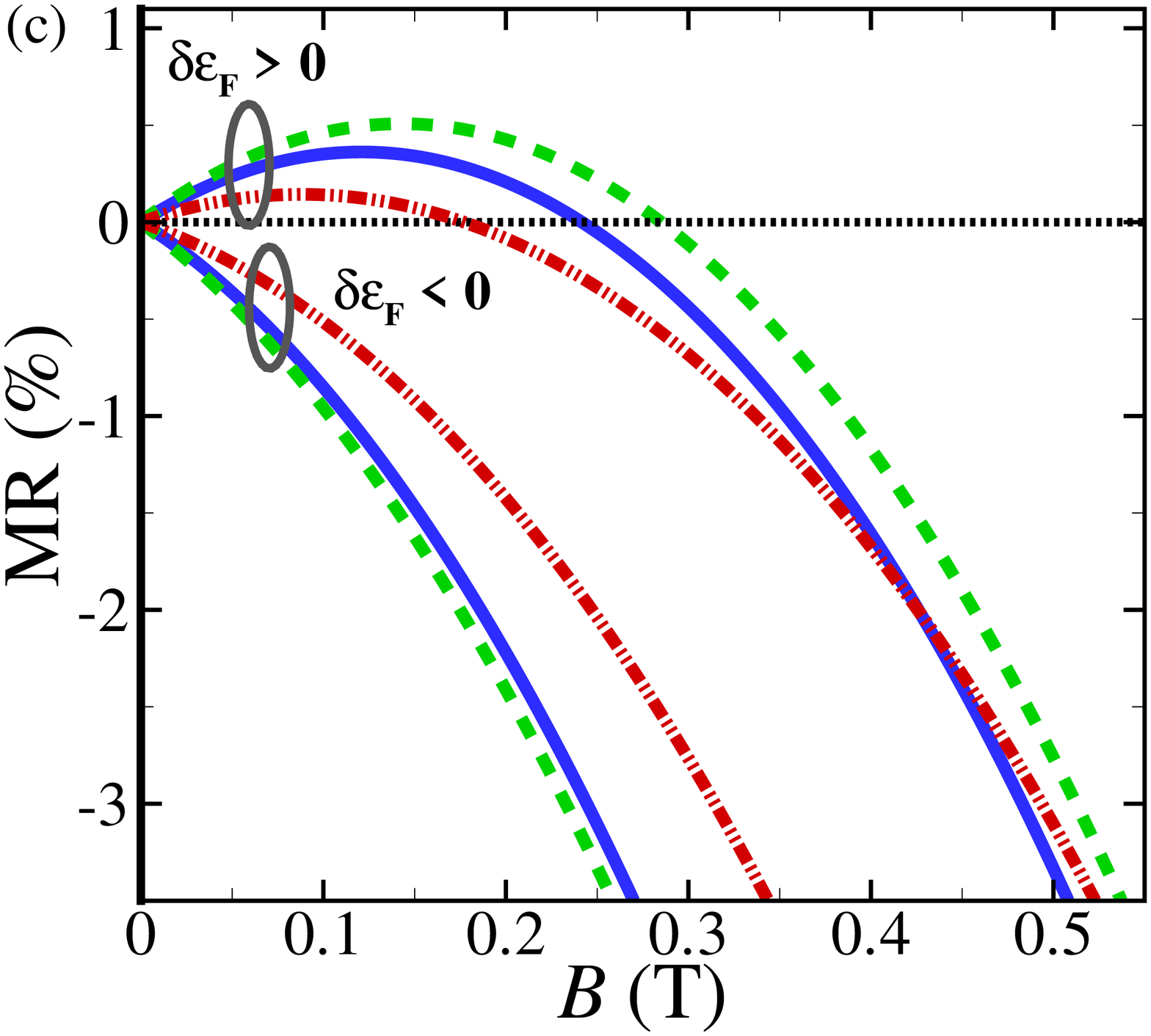}}    

\caption{(Color online) (a) The MR versus magnetic field, $B$, for different values of valley polarization: $\delta\varepsilon_{\rm F}=1\,\rm{meV}$ (solid red), $\delta\varepsilon_{\rm F}=2\,\rm{meV}$ (dashed blue), $\delta\varepsilon_{\rm F}=3\,\rm{meV}$ (dashed-dotted green) and $\delta\varepsilon_{\rm F}=4\,\rm{meV}$ (dashed-dotted-dotted pink). (b) The MR versus magnetic field, $B$, for different values of band gap: $\Delta = 10\,\rm{meV}$, (solid red), $\Delta = 20\,\rm{meV}$ (dashed blue), $ \Delta = 30\,\rm{meV}$ (dashed-dotted green). For $\delta\varepsilon_{\rm F}>0$, we set $\varepsilon_{{\rm F}_1}=1.1\Delta$ and $\varepsilon_{{\rm F}_2}=1.05\Delta$ and vice versa for $\delta\varepsilon_{\rm F}<0$. (c) The MR versus magnetic field, $B$, in the presence and absence of in-scattering and intervalley scattering terms. The color description of the curves is the same as Fig.~\ref{fig1}(a).}
    \label{fig4}
    \end{figure}

It is also interesting to note that increasing the difference in Fermi energies of the two valleys has a different impact on the system when $\delta\varepsilon_{\rm F}>0$ or $\delta\varepsilon_{\rm F}<0$. As illustrated in Fig. \ref{fig4}(a), while increasing $\delta\varepsilon_{\rm F}$ substantially affects the $\delta\varepsilon_{\rm F}>0$ case, the MR in the $\delta\varepsilon_{\rm F}<0$ case does not change considerably especially for smaller magnetic fields. Meanwhile in both cases, the increase in the polarization of the valleys has a positive impact on the MR of the system. We can also see the effect of the band gap on the MR in Fig. \ref{fig4}(b).  Clearly as the OMM is inversely proportional to the band gap, we can see that the MR in a small band-gap system is more substantially affected in the presence of the OMM.

On the other hand, similar to the unpolarized system, the intervalley scattering has the same negative contribution in MR for the $\delta\varepsilon_{\rm F}<0$ case; while on the contrary, for the $\delta\varepsilon_{\rm F}>0$ case the contribution of the intervalley scattering is positive (Fig. \ref{fig4}(c)). We should note that practically, in the valley-polarized case the intervalley scattering can not be very large, because it can destroy the valley polarization.
 
\section{Summary and discussions}

To summarize, we have revealed that going beyond the simple relaxation-time approximation in 2D multivalley systems can effectively modify the quantum transport properties such that the sign of the MR can be reversed for the high-density regime. In the case of the Fermi level close to the band bottom, the inclusion of the intervalley scatterings results in a significant drop of the MR when one of the valleys is depleted whose size increases as the intervalley scattering grows.

\begin{figure}[!ht]
  \captionsetup[subfigure]{labelformat=empty}
\centering
    \subfloat[]{\includegraphics[width=0.7\linewidth]{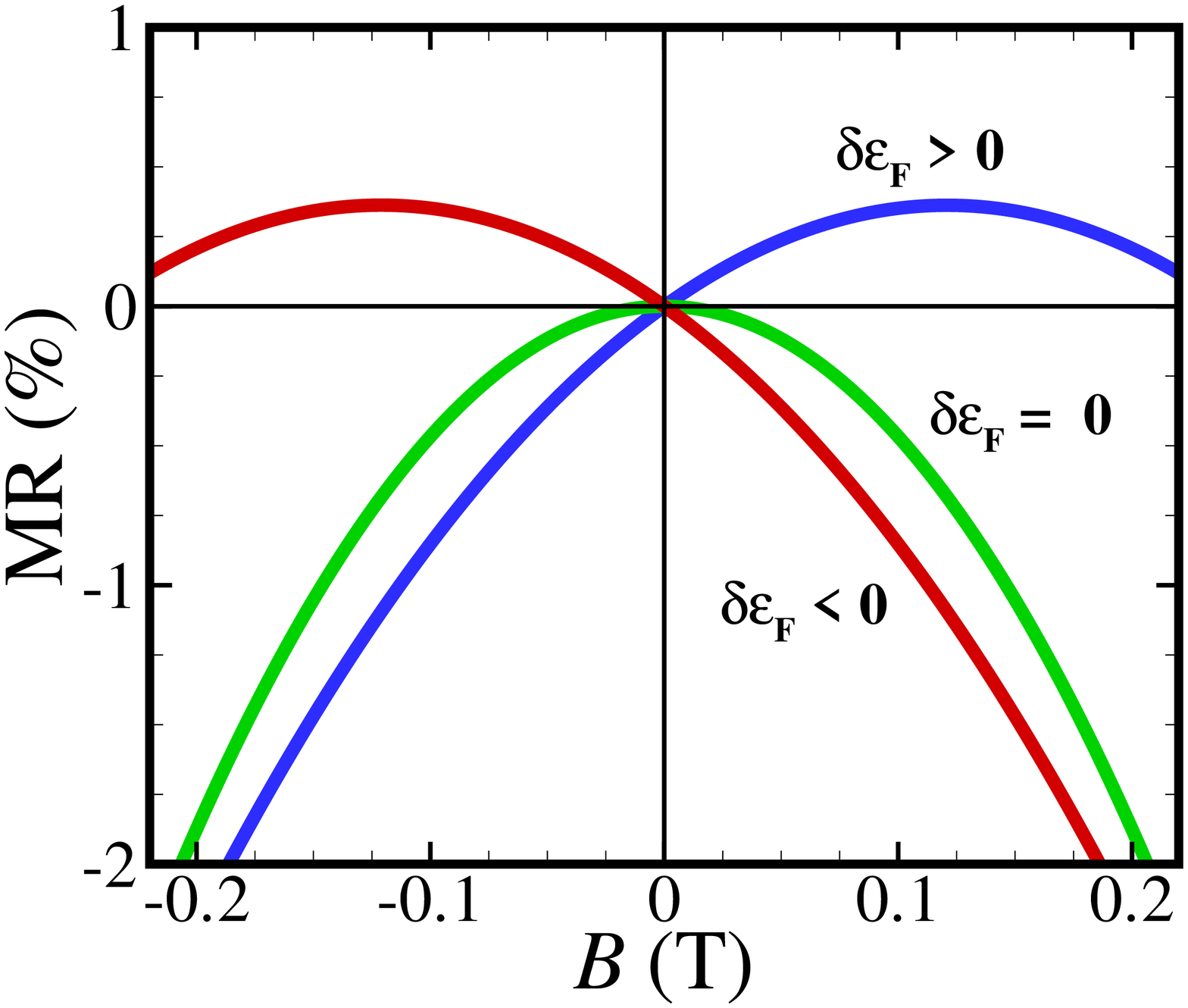}}
\caption{(Color online) The magnetic-field dependence of the MR in low magnetic fields for both time-reversal symmetric ($\delta\varepsilon_{\rm F}=0$, green curve) and asymmetric ($\delta\varepsilon_{\rm F}\neq0$ red and blue curves) cases. }
    \label{fig5}
    \end{figure}

Furthermore, we have studied the MR in a multivalley time-reversal-broken system in the presence of the OMM. If we assume the $\delta\varepsilon_{\rm F}>0$ ($\delta\varepsilon_{\rm F}<0$) as a positive (negative) polarization, we have demonstrated that the OMM induces an intrinsic Hall conductivity which is negative in any polarizations and its absolute value grows with the magnetic field. This is in contrast to the anomalous Hall conductivity which is positive (negative) for a positive (negative) polarization.
Figure. \ref{fig5} summarizes the MR results in both initially time-reversal-symmetric ($\delta\varepsilon_{\rm F}=0$) and time-reversal-broken ($\delta\varepsilon_{\rm F}>0$ and $\delta\varepsilon_{\rm F}<0$) systems. As expected the MR in a low-field limit behaves as $B^2$ for the time-reversal-symmetric case, while it shows a linear dependence on $B$ for the two time-reversal-broken curves. 

The completely different behaviors of the MR for positive and negative polarizations is also depicted here. While for the negative(positive) polarization and $B>0$ ($B<0$), the MR stays negative for all values of the magnetic field, it changes sign in the opposite case. This sign change in the MR in low magnetic fields is the direct consequence of the OMM consideration in the calculations. Our calculations also show that interestingly the sensitivity of $\delta\varepsilon_{\rm F}>0$ and $\delta\varepsilon_{\rm F}<0$ cases to the size of the polarization is completely different such that for $B>0$, increasing the polarization between two valleys significantly influences the MR in the former case, while it possesses a moderate effect on the latter. This polarization-dependent MR can be verified experimentally for example by illuminating the circularly polarized light with opposite senses in multivalley systems like TMDCs. We should note that since both Berry curvature and OMM are inversely proportional to the gap, the OMM-induced modifications are expected to be seen in systems with smaller band gaps.
 
We want to emphasize that in this paper, following an exact solution to the semiclassic Boltzmann equation, we focus on intrinsic mechanisms contributing to the magnetotransport involving the OMM and the Berry curvature induced anomalous velocity and the correction to the density of states. In Ref.~\cite{xiao2020linear}, it was indicated that the inclusion of the side jump contribution can also strengthen the intra-scattering contribution in the collision integral. Therefore, it is interesting to inspect the interplay between the extrinsic effects like side-jump and skew scattering and the intrinsic mechanisms described here in both 2D and 3D inversion-broken systems.  

Finally, we note that in this paper we have focused on semiclassical treatment of electron dynamics up to first order in electromagnetic fields. It has been shown that extending the semiclassical dynamics up to second order and the consequent magnetic field correction to the Berry curvature can give rise to a magnetononlinear anomalous Hall effect\cite{gao2019semiclassical,gao2014field}. In this case, in addition to Berry curvature and OMM, a new band geometric quantity, the anomalous orbital polarizability, plays a major role\cite{wang2022theory}. There is also a spin counterpart of anomalous orbital polarizability- the anomalous spin polarizability- which vanishes for the systems without spin-orbit coupling as in here. 
    
\section{Acknowledgement}
A. F acknowledges the support from Iran Science Elites Federation(ISEF).
\appendix
\section{}\label{ap-one}
Applying $\delta(\tilde{\varepsilon}_{k'}^{\tau'}-\tilde{\varepsilon}_{k}^{\tau})=\frac{\delta(k'-k'_0)}{\vert \partial(\tilde{\varepsilon}_{k'}^{\tau'}-\tilde{\varepsilon}_{k}^{\tau})/\partial k'\vert_{k'_0} }$ to  Eqs.~(\ref{G})-(\ref{I}), the integrals are easily found as
 \begin{align}
&G^{\tau}(k)=\frac{n_i}{4\pi\hbar^2}D_k^{\tau}\sum_{\tau'}\bigl[\vert V^{{\tau}{\tau}'}\vert^2(\frac{k'_0D_{k'_0}^{\tau'}}{\tilde{v}_{k'_0}^{\tau'}})(1+\frac{\Delta^2}{\varepsilon(k)\varepsilon(k'_0)})\bigr],\\ 
&H^{\tau\tau'}(k)=\frac{n_i}{4\pi\hbar^2}\vert V^{{\tau}{\tau}'}\vert^2k'_0(\frac{D_{k}^{\tau}D_{k'_0}^{\tau'}}{\tilde{v}_{k}^{\tau}})(1+\frac{\Delta^2}{\varepsilon(k)\varepsilon(k'_0)}),\\
&I^{\tau\tau'}(k)=\frac{n_i}{4\pi\hbar^2}\vert V^{{\tau}{\tau}'}\vert^2k'_0(\frac{D_{k}^{\tau}D_{k'_0}^{\tau'}}{\tilde{v}_{k}^{\tau}})(\frac{\hbar^2v_F^2 kk'_0}{\varepsilon(k)\varepsilon(k'_0)}),
\end{align} 
where $k'_0$ is the root of $\tilde{\varepsilon}_{k'}^{\tau'}-\tilde{\varepsilon}_{k}^{\tau}=0$.

\section{}\label{ap-two}
To find $a^{\tau}(\phi)$ (the same procedure is applied for $b^{\tau}(\phi)$), we use the Fourier series as
 \begin{equation}\label{a}
 \begin{split}
a^{\tau}(\phi)&=a_0^{\tau}+a_{c1}^{\tau}\cos\phi+a_{c2}^{\tau}\cos 2\phi+\dots\\
&+a_{s1}^{\tau}\sin\phi+a_{s2}^{\tau}\sin 2\phi+\dots,
\end{split}
\end{equation} 
and the derivative of the form 
 \begin{equation}\label{aa}
 \begin{split}
\partial_{\phi}a^{\tau}(\phi)&=-a_{c1}^{\tau}\sin\phi-2a_{c2}^{\tau}\sin 2\phi-\dots\\
&+a_{s1}^{\tau}\cos\phi+2a_{s2}^{\tau}\cos 2\phi+\dots.
\end{split}
\end{equation} 
Plugging \eqref{a} and \eqref{aa} in \eqref{B3} and equating the coefficients of $\sin\phi$ and $\cos\phi$, we will see that the only nonzero coefficients in Fourier series~\eqref{a} are $a_0^{\tau}$, $a_{c1}^{\tau}$ and $a_{s1}^{\tau}$ as expected (since the Fermi surface is isotropic and the scattering is uniform) which are found solving the following equations
 \begin{align}
&G^{\tau}(k)a_0^{\tau}=2\pi[H^{\tau\tau}(k)a_0^{\tau}+H^{\tau(-\tau)}(k)a_0^{-\tau}],\label{1}\\
&F^{\tau}(k)a_{c1}^{\tau}=G^{\tau}(k)a_{s1}^{\tau}-\pi[I^{\tau\tau}(k)a_{s1}^{\tau}+I^{\tau(-\tau)}(k)a_{s1}^{-\tau}],\label{2}\\
&1-F^{\tau}(k)a_{s1}^{\tau}=G^{\tau}(k)a_{c1}^{\tau}-\pi[I^{\tau\tau}(k)a_{c1}^{\tau}-I^{\tau(-\tau)}(k)a_{c1}^{-\tau}].\label{3}
\end{align} 
Putting $\tau=\pm 1$ in the above equations, we will have four coupled equations for $a_{c1}^{+}$, $a_{s1}^{+}$, $a_{c1}^{-}$ and $a_{s1}^{-}$ (Eqs.~\eqref{2} and \eqref{3}) and $a_0^{+}$ and $a_0^{-}$ can be found from Eq.~\eqref{1} plus the condition for conservation of the electron number $\sum_{k,{\tau}}g_k^{\tau}=0$. Having found these unknown quantities, we will have
 \begin{equation}\label{a2}
a^{\tau}(\phi)=a_0^{\tau}+a_{c1}^{\tau}\cos\phi+a_{s1}^{\tau}\sin\phi.
\end{equation} 
Following the same procedure, $b^{\tau}(\phi)$ is given by
 \begin{equation}\label{b2}
b^{\tau}(\phi)=b_0^{\tau}+b_{c1}^{\tau}\cos\phi+b_{s1}^{\tau}\sin\phi,
\end{equation} 
where $b_0^{\tau}$, $b_{c1}^{\tau}$ and $b_{s1}^{\tau}$ are found solving the following equations
 \begin{align}
&G^{\tau}(k)b_0^{\tau}=2\pi[H^{\tau\tau}(k)b_0^{\tau}+H^{\tau(-\tau)}(k)b_0^{-\tau}],\label{12}\\
&1+F^{\tau}(k)b_{c1}^{\tau}=G^{\tau}(k)b_{s1}^{\tau}-\pi[I^{\tau\tau}(k)b_{s1}^{\tau}+I^{\tau(-\tau)}(k)b_{s1}^{-\tau}],\label{22}\\
&-F^{\tau}(k)b_{s1}^{\tau}=G^{\tau}(k)b_{c1}^{\tau}-\pi[I^{\tau\tau}(k)b_{c1}^{\tau}-I^{\tau(-\tau)}(k)b_{c1}^{-\tau}].\label{32}
\end{align} 
Note that we do not need to find $a_0^{\tau}$ and $b_0^{\tau}$, because they are the constant parts of the distribution function and have no contribution in transport integrals.

\end{document}